\documentclass[conference]{IEEEtran}
\IEEEoverridecommandlockouts

\usepackage[utf8x]{inputenc}

\usepackage{amssymb}
\usepackage{amsmath}
\usepackage{pgfplots}
\usepackage{multirow}
\usepackage{comment}
\usepackage{booktabs}
\usepackage{comment}
\usepackage{flushend}
\usepackage{adjustbox}
\usepackage{bm}

\usepackage{subfigure}

\begin{document}
\pgfplotsset{ymin=0}

\title{AM-MobileNet1D: A Portable Model\\for Speaker Recognition
}

\author{\IEEEauthorblockN{
Jo\~ao Ant\^onio Chagas Nunes\,$^{1}$, 
David Macêdo\,$^{1,2}$, and
Cleber Zanchettin\,$^{1,3}$
}
\IEEEauthorblockA{\,
$^{1}$Centro de Inform\'atica, Universidade Federal de Pernambuco, Recife, Brasil\\
$^{2}$Montreal Institute for Learning Algorithms, University of Montreal, Quebec, Canada\\
$^{3}$Department of Chemical and Biological Engineering, Northwestern University, Evanston, United States of America\\
Emails: \{jacn2, dlm, cz\}@cin.ufpe.br}
}

\maketitle

\begin{abstract}
Speaker Recognition and Speaker Identification are challenging tasks with essential applications such as automation, authentication, and security. Deep learning approaches like SincNet and AM-SincNet presented great results on these tasks. The promising performance took these models to real-world applications that becoming fundamentally end-user driven and mostly mobile. The mobile computation requires applications with reduced storage size, non-processing and memory intensive and efficient energy-consuming. The deep learning approaches, in contrast, usually are energy expensive, demanding storage, processing power, and memory. To address this demand, we propose a portable model called Additive Margin MobileNet1D (AM-MobileNet1D) to Speaker Identification on mobile devices. We evaluated the proposed approach on TIMIT and MIT datasets obtaining equivalent or better performances concerning the baseline methods.  Additionally, the proposed model takes only 11.6 megabytes on disk storage against 91.2 from SincNet and AM-SincNet architectures, making the model seven times faster, with eight times fewer parameters. 

\end{abstract}


\IEEEpeerreviewmaketitle

\section{Introduction}
\label{introduction}

\begin{figure}%
\centering
\subfigure[Speaker Identification]{
\includegraphics[width=0.8\columnwidth]{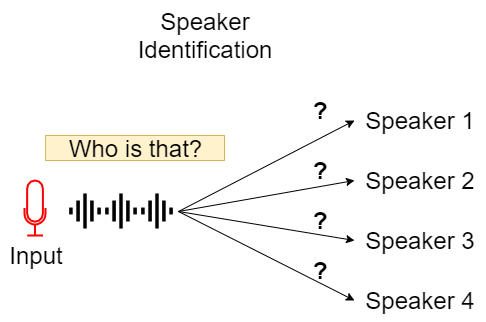}
}
\hspace{1cm}
\subfigure[Speaker Verification]{
\includegraphics[width=0.8\columnwidth]{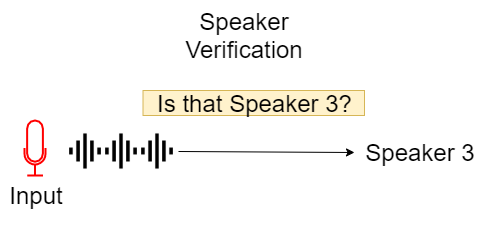}
}

\caption{Difference between Speaker Identification and Verification.}
\label{fig:verification_vs_identification}
\end{figure}

\begin{figure*}[!t]
\centering
\includegraphics[width=0.9\textwidth]{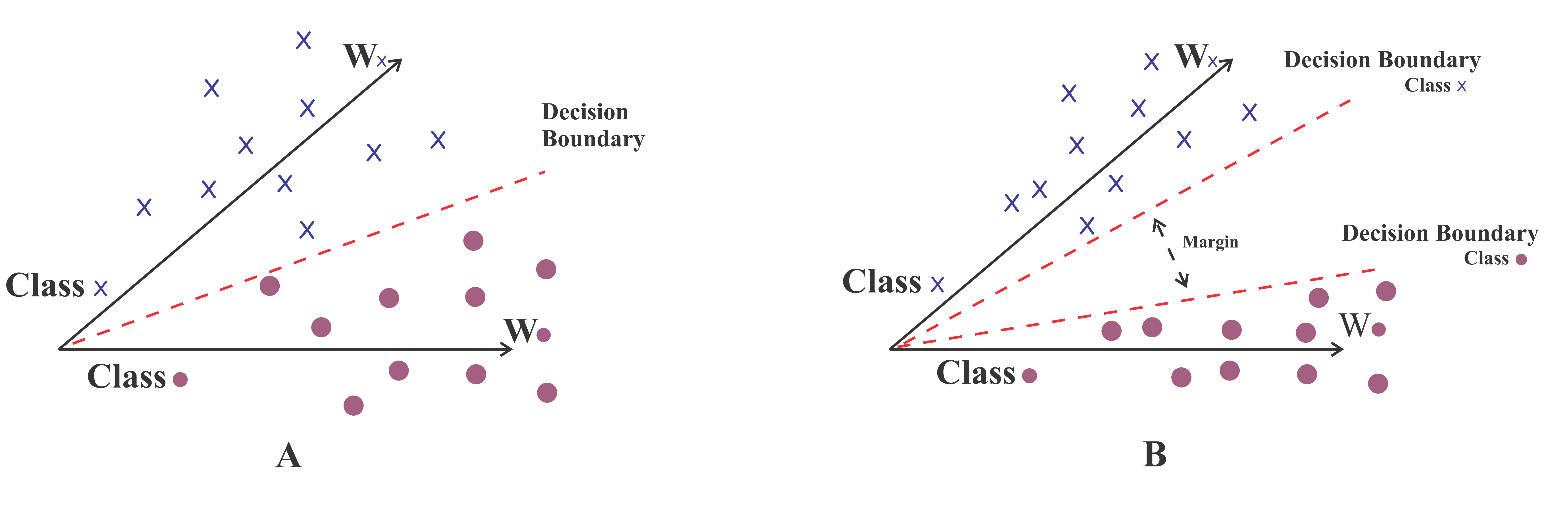}
\caption{Comparison between Softmax (A) and AM-Softmax (B). Adapted from  \cite{AM-Softmax}.}
\label{fig:SoftmaxVSAMSoftmax}
\end{figure*}

Speaker Recognition is a very active area with applications in biometric authentication, identification, security, among others \cite{Beigi}. This area has two principal tasks: Speaker Identification and Speaker Verification. Figure \ref{fig:verification_vs_identification} shows a representation of these two tasks. Speaker Identification (Figure \ref{fig:verification_vs_identification} (a)) is the task of identifying who spoke something among a determined list of speakers. Whereas, Speaker Verification (Figure \ref{fig:verification_vs_identification} (b)) aims to classify whether an audio sample belongs to a pre determined person or not. Before the Deep Learning models, most of the literature techniques to tackle this problem were based on $i$-vectors methods \cite{Dehak10frontend, i-vector, i-vector2}. These methods try to find patterns on audio signals and classify them using techniques such as  Probabilistic Linear Discriminant Analysis (PLDA) \cite{Prince2007ProbabilisticLD}, heavy-tailed PLDA \cite{htPLDA}, and  Gaussian PLDA \cite{GaussPLDA}.

The Speaker Recognition task is challenging because of the high dimensionality found on audio signals, which are complex to be modeled in low and high-level features. These features need to be good enough to distinguish different speakers. Over the years, pattern recognition, image processing, and signal processing tasks, are being successfully solved by Deep Neural approaches \cite{deep2, deep3, SwishNet}. Deep learning based models learn complex patterns that may not be spotted by humans or even traditional machine learning approaches. These methods are more complex but usually produce better results than conventional approaches. Indeed, Convolutional Neural Networks (CNN) \cite{CNNs-LeCun} have already proved to be the best choice for image classification, detection, and recognition tasks \cite{deep_image, deep_image2, deep_image3}.

An exciting Deep Learning-based model to Speaker Recognition is the SincNet model \cite{SincNet}. It is a hybrid model, like \cite{hybrid1, hybrid2, hybrid3, hybrid4, hybrid5}, that uses Sinc convolutional layers in its first layer. The Sinc convolutional layer uses parameterized Sinc functions to convolve the waveform of the audio signal and uses this information to Speaker Identification. A more efficient approach to the problem is the AM-SincNet \cite{AMSincNet}, which is a combination of the SincNet and the AM-Softmax loss function \cite{AM-Softmax}. The AM-SincNet solves the problem of the traditional SincNet by replacing the Softmax layer with an improved AM-Softmax layer. In fact, this change was enough to show an improvement of approximately $40\%$ in the Frame Error Rate when compared to the traditional SincNet. 

These models are being applied to real-world purposes to aggregate value, performance, and usability to end-user applications. However, end-user applications are becoming more mobile. The mobile devices revolutionized many aspects of how we live our lives. From text messaging, to map apps for localization, and always having a camera to document what we see and experience, the smartphones, and other portable computations are incredibly useful devices. As the massive use of these dispositive, more server applications are migrating to the client device for better usability, performance, and quality. 

The problem with Deep Learning models in this context is that they are not mobile; they are often large size models that require substantial computational effort, memory, storage size, and energy to be executed. Most of them need dedicated hardware, like Graphics Processing Units (GPUs), to have a reasonable inference time. %

Some deep learning approaches, as the MobileNet model \cite{mobilenet}, are designed to work on mobile devices. The MobileNet architecture introduces the concept of depthwise separable convolutions that enables it to be lighter and faster than the traditional deep learning models. The depthwise separable convolutions break the convolution process into two parts: the depthwise and the pointwise operation. It impacts on reducing the number of parameters on the convolutional layer and improving the time to proceed a convolution. %

In this paper, we proposed two new deep learning models based on the MobileNetV2 \cite{mobilenetv2} and the AM-Softmax \cite{AM-Softmax} loss function to deal directly with raw audio signals for Speaker Identification task. The proposed models are MobileNet1D, and AM-MobileNet1D, both of them evaluated against the SincNet and AM-SincNet models on the TIMIT \cite{timit} and MIT \cite{mit} datasets. The experiments consider the Frame Error Rate (FER), Classification Error Rate (CER), the inference time, the number of parameters, and the model size. On the tested scenarios, the proposed methods show to be competitive, getting equally good, or even better results than the SincNet and AM-SincNet. In addition, the proposed methods are about $7$ times faster, have $8$ times fewer parameters, and take up to $8$ times less space on disk than the SincNet and the AM-SincNet.

The following sections are organized as: Section \ref{related_work}, we present the background of this paper, the proposed methods are introduced in Section \ref{proprosed_methods}, Section \ref{experiments} explains how the experiments were built, the results are discussed at Section \ref{results}, and finally at Section \ref{conclusion} we made our conclusions.

\section{Background}
\label{related_work}

\begin{figure}[!b]
\centering
\includegraphics[width=0.9\columnwidth]{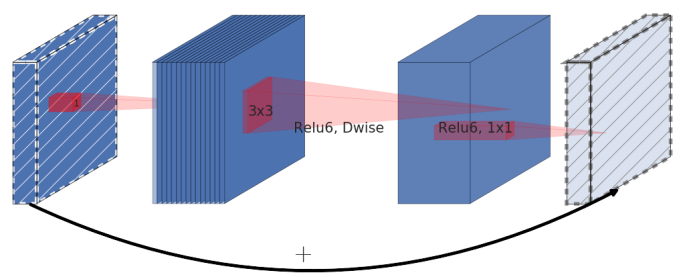}
\caption{MobileNetV2 Inverted Residual Block. Adapted from \cite{mobilenetv2}.}
\label{fig:inverted_residual_block}
\end{figure}

\begin{figure*}%
\centering
\subfigure[MobileNetV2]{
\includegraphics[width=0.35\columnwidth]{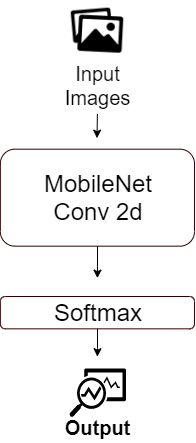}
}
\hspace{1.5cm}
\subfigure[MobileNet1D]{
\includegraphics[width=0.35\columnwidth]{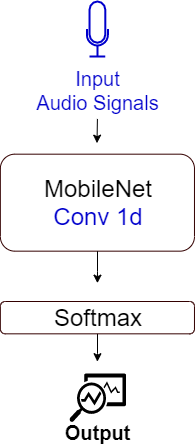}
}
\hspace{1.5cm}
\subfigure[AM-MobileNet1D]{
\includegraphics[width=0.35\columnwidth]{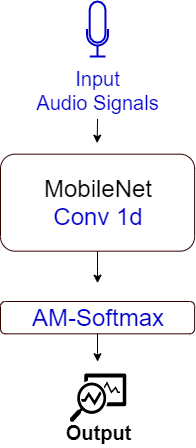}
}
\caption{MobileNet and the proposed architectures.}
\label{fig:models_evolution}
\end{figure*}

Deep learning models usually use the Softmax loss function on its final layer to map the extracted features into different classes. It works by delimiting a linear surface, which is used as a decision boundary to split the samples into their respective categories. The problem with the Softmax decision boundary is that it does not maximize the distance of samples from different classes, which may cause misclassification of samples that are placed too close to the linear separation surface. This characteristic may spoil the Softmax efficiency on tasks that require to measure the distance between samples to make a decision. A new loss function based on the Softmax was proposed by \cite{AM-Softmax} to solve this problem, it is called Additive Margin Softmax (AM-Softmax) and introduces an additive margin to the Softmax decision boundary. The additive margin forces samples from the same class to lay closer to each other, while maximizes the distance of samples from different classes.

The Additive Margin Softmax (AM-Softmax), proposed by \cite{AM-Softmax} to deal with face recognition problems, is a based Softmax loss function that introduces an additive margin of separation to the boundaries of each class. It means that each linear surface of separation has an additional region (the additive margin) that may not be used for any sample, a safe zone. An implication of the additive margin used on the AM-Softmax is that it maximizes the distance between samples from different classes, while at the same time, it forces the samples from the same class to be closer to each other. Figure \ref{fig:SoftmaxVSAMSoftmax} shows the behavior of the samples when the AM-Softmax is used comparing traditional Softmax (A) and the AM-Softmax (B). The AM-Softmax added two more parameters in comparison to traditional Softmax. The new parameter $s$ scale the samples and parameter $m$ controls the size of the additive margin. The AM-Softmax equation is written as:

\begin{equation}
\label{eq:AM-Softmax}
Loss = -\frac{1}{n} \sum^n_{i=1}log \frac{\phi_i}{\phi_i + \sum^c_{j=1, j \neq y_i }exp(s(W_{j}^T f_i))}
\end{equation}
\begin{equation}
\phi_i = exp(s(W_{y_i}^T f_i-m))
\end{equation}

where $W$ is the weight matrix, and $f_i$ is the input from the $i$-th sample for the last fully connected layer. The $W_{y_i}^T f_i$ is also known as the target logit for the $i$-th sample. The $s$ and $m$ are the parameters responsible for scaling and additive margin, respectively.

Besides being applied to face recognition, the AM-Softmax was also applied to speaker recognition by \cite{AMSincNet}. The AM-SincNet improves the results of the SincNet \cite{SincNet}. The SincNet overcomes traditional convolutional networks on the mentioned task. However, it uses the Softmax loss function on its last layer. According to \cite{AMSincNet}, the Softmax loss function was spoiling the SincNet results, so the authors proposed the AM-SincNet model. The AM-SincNet model is based both in the SincNet architecture and the AM-Softmax loss function. The AM-SincNet improves the SincNet results by replacing a traditional Softmax layer to an improved AM-Softmax layer. The method was evaluated on TIMIT \cite{timit} dataset and reduced the Frame Error Rate in about $40\%$ compared to the traditional SincNet. Despite the good results on the speaker identification task, if we consider the mobile application of the model, the SincNet is not appropriated. The model needs too much computational effort to make predictions in a reasonable inference time.

A lightweight and fast deep learning model is the MobileNetV2 \cite{mobilenetv2}, which is a new version of the MobileNet \cite{mobilenet}, both of them were built to work with images in classification, detection, and segmentation tasks. MobileNetV2 is a reasonable choice when applications need to run on mobile devices that have limited computational resources and cannot support large size models. In comparison to the MobileNet, the MobileNetV2 introduces the Inverted Residual Block, shown in Figure \ref{fig:inverted_residual_block}, where the thickness of each block indicates how much channels it has. The Inverted Residual Block connects the bottlenecks layers reducing the amount of memory needed in comparison to the MobileNet Residual Block that connects layers with higher number of channels. What makes the MobileNet architecture unique is the depthwise separable convolutions, which is also part of the MobileNetV2 model. The depthwise separable convolution is a new way to make time-efficient convolutions, and it is built by splitting the traditional process of convolution into two other operations: depthwise convolution and pointwise convolution. They work by sharing the convolution effort between them. The depthwise convolution applies a single convolution filter on each channel of the input sample, while the pointwise convolution uses a $1\times1$ convolution filter on the output of the depthwise operation. Thus, the pointwise work is to combine the outputs from the depthwise operation into the $n$ required depthwise separable convolutional filters. This modification drastically reduces the computational effort needed to process the data and the model size, while speeds up the entire process when compared to traditional convolutions. According to \cite{mobilenet}, the use of the depthwise separable convolutions of size $3\times3$ helps to reduce the computational effort of a model in $8$ to $9$ times when compared with traditional convolutional layers.

Another advantage of the MobileNet and MobileNetV2 is that they principal operation, the depthwise separable convolutions, can be built in any modern deep learning framework because the depthwise and the pointwise convolutions are made from traditional convolution operations. According to \cite{mobilenetv2}, the MobileNetV2 overcomes, in both accuracy and model complexity, state-of-the-art real-time detectors on COCO dataset \cite{coco} for the object detection task. Combined with SSDLite detection module \cite{mobilenetv2}, which is a modified version of the Single Shot Detector (SSD) \cite{ssd}, it also consumes $20$ times less computational effort and has $10$ times fewer parameters than the YOLOv2 \cite{yolov2}.

Yet, the MobileNets are designed to computer vision tasks, so their operations, including the depthwise separable convolutions, are built to work with images. Thus, to work directly with audio signals in the speaker recognition problem, these models need some adaptation.

\begin{table*}[!t]
\renewcommand{\arraystretch}{1.5}
\centering
\caption{ Comparison of the speaker recognition models on the TIMIT and MIT dataset using the metrics Frame Error Rate (FER), Classification Error Rate (CER), Inference Time, Number of Parameters, and Model Size.}
\label{tab:results_error}
\resizebox{\linewidth}{!}{%
\begin{tabular}{lcccccccc}
\toprule
& \multicolumn{2}{c}{TIMIT} & & \multicolumn{2}{c}{MIT} & Inference Time & Parameters & Model Size \\ \cmidrule{2-3} \cmidrule{5-6}
&                   FER (\%)[$\downarrow$]     & CER (\%)[$\downarrow$] &    & FER (\%)[$\downarrow$]   & CER (\%)[$\downarrow$]  & (ms)[$\downarrow$] & (total)[$\downarrow$] & (mb)[$\downarrow$]\\ \midrule
SincNet             & 44.64 & 1.15 &    & 34.65 & 1.27  &  42.60\:$\pm$\:1.5766 & 22,787,134 & 91.2\\ %
AM-SincNet          & 27.86 & 0.50 &    & \textbf{30.30} & \textbf{0.52}  & 41.17\:$\pm$\:0.7529  & 22,787,134 & 91.2\\ %
MobileNet1D         & 26.50 & 0.57 &    & 38.75 & 2.54   & \textbf{5.88\:$\pm$\:0.1584} & \textbf{2,825,942} & \textbf{11.6} \\ %
AM-MobileNet1D      & \textbf{21.30} & \textbf{0.43} &    & 35.55 & 2.19  & \textbf{5.85\:$\pm$\:0.1673} &  \textbf{2,825,942} & \textbf{11.6} \\
\bottomrule
\end{tabular}
}
\end{table*}

\section{Proposed Models}
\label{proprosed_methods}

In this section, we propose the lightweight deep learning models MobileNet1D and AM-MobileNet1D to tackle the speaker recognition problem, especially the speaker identification tasks. First, we present the adaptation of the MobileNetV2 for speaker recognition (called MobileNet1D). Then we describe how the modified MobileNet1D built the AM-MobileNet1D for the same task.

\subsection{MobileNet1D}
\label{mobilenet1d}

The MobileNet \cite{mobilenet} is a convolutional neural network built to be fast and light, which makes it appropriate to work with high dimensional data like audio signals. Although, the MobileNet architecture is modeled to work with two-dimensional structs like images. However, unlike images, the audio signals are modeled in one-dimensional. Indeed, it is necessary to adapt the MobileNet architecture to fit audio signals. Thus, to tackle the speaker recognition problem, we made some adaptation on the MobileNetV2 \cite{mobilenetv2} architecture.

Nevertheless, to adapt the MobileNetV2 architecture, we first go to its core, the depthwise separable convolution. Every element on the depthwise separable convolution is designed to work with two-dimensional structs, the convolutions, the batch normalization, and the pooling layers. For this reason, it is necessary to adapt these basic operations on the depthwise and pointwise layers. Thus, we first change the convolutions on the depthwise struct to proceed 1d convolutions instead of 2d. Batch normalization and pooling operations were also replaced from 2d to 1d implementations. These modifications not only prepared MobileNet to process audio signals, but it also speeds up the network by reducing its complexity and model size. Figure \ref{fig:models_evolution} shows an illustration of the MobileNet with the proposed MobileNet1D and AM-MobileNet1D methods. In Figure \ref{fig:models_evolution} (b), the audio signals are fed into the MobileNet architecture adapted with 1d depthwise separable convolutions, then the extracted features feed a Softmax layer which will output the probabilities of the sample belong to each class.

\subsection{AM-MobileNet1D}
\label{am-mobilenet1d}

As presented in Figure \ref{fig:models_evolution} (b), the proposed method MobileNet1D uses a Softmax loss function to map the extracted features into their respective classes. However, the Softmax loss function may not be the best choice for this type of problem. Indeed, the Additive Margin Softmax (AM-Softmax) loss function has shown to be a more sophisticated approach for classification problems. It let the models achieve better results not only in the face recognition problem where it was first proposed but also in the speaker recognition problem, as shown in \cite{AMSincNet}. The AM-Softmax has improved significantly the Frame Error Rate measured in the AM-SincNet in comparison to the calculated in the traditional SincNet.

Nevertheless, the AM-Softmax introduces two new parameters when compared to the traditional Softmax. They are the $s$ parameter which is responsible for scaling the features and the $m$ parameter that controls the size of the additive margin. In \cite{AM-Softmax}, where the AM-Softmax loss function was proposed, it was established that the scaling factor $s$ needs to be high enough to work well. It also can be learned by the training process of the network. However, it would make the network convergence too slow. Thus, \cite{AM-Softmax} states that a fixed value of $s$ works just fine and it proposes to set $s=30$. On the other hand, the margin size of $m$ may vary in the range $0\!\leq\!m\!\leq\!1$. In fact, \cite{AMSincNet} made several experiments on the dataset TIMIT to see the influence of the margin parameter on the Frame Error Rate (FER), it concludes that at the end of training every $m$ value tested in the range $0,35\!\leq\!m\!\leq\!0,80$ converges the FER to the same place. Also, none of the $m$ values tested on AM-SincNet got higher (worse) error rates than the SincNet using the Softmax loss function.

So, the proposed Additive Margin MobileNet1D (AM-MobileNet1D) is a combination of the MobileNet1D described in the last subsection and the AM-Softmax loss function. It replaces the last Softmax layer of the MobileNet1D for an improved Additive Margin Softmax layer. With this adaptation, the AM-MobileNet1D gets two new parameters from the AM-Softmax function, the scaling parameter $s$, and the margin size $m$. Following \cite{AM-Softmax}, we set $s=30$ for the proposed model. In addition, for the margin size parameter, as \cite{AMSincNet} stated that every tested value of $m$ converged the FER to the same place, we fixed $m=0.5$ in our experiments. Figure \ref{fig:models_evolution} (c) shows an illustration of the proposed AM-MobileNet1D where the input audio signals fed a MobileNet1D and then the extracted features are mapped into their classes by an AM-Softmax loss function.

\section{Experiments}
\label{experiments}

In this section, we describe the experimental protocol to evaluate the proposed models, MobileNet1D and AM-MobileNet1D, considering the baseline methods SincNet and AM-SincNet on the Speaker Identification task. %

\subsection{Datasets}
\label{datasets}

We use two different speaker datasets. The first one is the well-know TIMIT \cite{timit} dataset, which contains audio samples from speakers of the eight main American dialects. It contains $630$ different speakers, and each one of them was instructed to read $10$ phonetically rich sentences. The dataset is actually divided into two distinct sets, the training set and the testing set. To a fear comparison, we follow the same protocol as \cite{SincNet} and \cite{AMSincNet}, where the experiments use only the training set of the TIMIT dataset, which contains $462$ speakers and $3,969$ audio samples.

The second dataset used to evaluate the models is the MIT \cite{mit} dataset, which is built from audio samples collected from mobile devices in a variety of environments in order to induce the presence of noise. Like the TIMIT dataset, the MIT dataset is also divided into two distinct sets, the set of authentic users and the set of imposters. The data were collected in two separate sessions for the authentic users and one more session for the imposters. Each session lasted for $20$ minutes and was recorded on different days. In order to simulate normal daily conditions, the sessions were recorded in three different places: a quiet office, a mildly noisy lobby, and a busy street intersection. The MIT dataset contains audio samples from $88$ different speakers, $39$ women, and $49$ men. Analyzing the imposters' set of the MIT dataset, we saw that not every authentic user has its own imposter. So, to avoid the imbalance problem, we only use the authentic users set, as the imposters set is not complete. The authentic users set contains $5,184$ audio samples from $48$ different speakers, where each speaker recorded exactly $108$ samples.

Each dataset was divided into two different sets, the training set, and the testing set. For the TIMIT dataset, five utterances of each speaker were used to train the network, while the remaining three were used to test. On the other hand, for the MIT dataset, $70\%$ of the samples from each speakers were selected for the training process, while the remaining samples were  used for testing. In both datasets, the samples that compose the training and the testing set were chosen randomly. Yet, in each dataset, the number of samples was balanced between the classes. 

\begin{figure*}%
\centering
\subfigure[TIMIT]{
\begin{tikzpicture}
\begin{axis}[
width=0.95\columnwidth,
axis lines = left,
xlabel = Epochs,
ylabel = {FER ($\%$)},
]
\addplot [
    color=red,
    line width=0.55mm,
]coordinates {
(0, 97.25)
(8, 68.20)
(16, 55.32)
(24, 52.84)
(32, 50.29)
(40, 49.58)
(48, 46.67)
(56, 45.20)
(64, 45.40)
(72, 50.08)
(80, 43.49)
(88, 44.62)
(96, 44.83)
(104, 46.93)
(112, 47.10)
(120, 43.88)
(128, 45.71)
(136, 46.52)
(144, 44.89)
(152, 43.70)
(160, 46.54)
(168, 42.53)
(176, 43.56)
(184, 43.56)
(192, 46.83)
(200, 43.69)
(208, 43.44)
(216, 43.41)
(224, 42.55)
(232, 42.81)
(240, 44.19)
(248, 44.27)
(256, 45.32)
(264, 45.22)
(272, 43.64)
(280, 42.69)
(288, 43.64)
(296, 43.84)
(304, 44.37)
(312, 42.07)
(320, 46.39)
(328, 43.70)
(336, 47.96)
(344, 44.99)
(352, 44.64)
};
\addlegendentry{SincNet}
\addplot [
color=red,
dashed,
line width=0.55mm,
]
coordinates {
(0, 99.06)
(8, 68.03)
(16, 58.37)
(24, 50.30)
(32, 43.46)
(40, 50.24)
(48, 40.54)
(56, 40.16)
(64, 38.02)
(72, 37.50)
(80, 34.89)
(88, 33.80)
(96, 33.68)
(104, 37.59)
(112, 32.42)
(120, 31.65)
(128, 30.49)
(136, 31.84)
(144, 30.92)
(152, 29.85)
(160, 30.65)
(168, 30.74)
(176, 30.11)
(184, 31.08)
(192, 30.73)
(200, 29.22)
(208, 29.20)
(216, 28.92)
(224, 28.16)
(232, 28.41)
(240, 28.62)
(248, 28.00)
(256, 28.64)
(264, 29.78)
(272, 28.32)
(280, 28.02)
(288, 28.41)
(296, 27.66)
(304, 28.54)
(312, 27.43)
(320, 27.37)
(328, 27.20)
(336, 27.42)
(344, 28.32)
(352, 27.86)
};
\addlegendentry{AM-SincNet}
\addplot [
    color=blue,
    line width=0.55mm,
]coordinates {
(0, 89.64)
(8, 49.59)
(16, 51.08)
(24, 39.68)
(32, 35.64)
(40, 35.10)
(48, 40.56)
(56, 33.48)
(64, 36.83)
(72, 32.03)
(80, 30.08)
(88, 33.82)
(96, 30.65)
(104, 29.50)
(112, 29.93)
(120, 29.63)
(128, 29.32)
(136, 28.87)
(144, 29.48)
(152, 31.22)
(160, 28.58)
(168, 30.10)
(176, 28.64)
(184, 37.62)
(192, 27.89)
(200, 28.33)
(208, 27.24)
(216, 29.21)
(224, 28.27)
(232, 27.16)
(240, 30.30)
(248, 27.78)
(256, 27.18)
(264, 26.50)
(272, 29.90)
(280, 28.55)
(288, 28.41)
(296, 28.81)
(304, 27.96)
(312, 28.91)
(320, 34.51)
(328, 27.31)
(336, 27.15)
(344, 26.86)
(352, 28.88)
};
\addlegendentry{MobileNet1D}

\addplot [
    color=blue,
    dashed,
    line width=0.55mm,
]coordinates {
(0, 89.02)
(8, 50.46)
(16, 37.60)
(24, 34.36)
(32, 31.44)
(40, 32.29)
(48, 33.30)
(56, 28.77)
(64, 26.80)
(72, 26.82)
(80, 26.55)
(88, 27.88)
(96, 27.10)
(104, 25.34)
(112, 24.87)
(120, 26.92)
(128, 25.26)
(136, 25.43)
(144, 23.38)
(152, 23.72)
(160, 23.39)
(168, 24.60)
(176, 22.80)
(184, 23.67)
(192, 22.57)
(200, 23.62)
(208, 22.63)
(216, 22.43)
(224, 22.75)
(232, 24.71)
(240, 23.25)
(248, 22.81)
(256, 22.38)
(264, 22.81)
(272, 22.30)
(280, 24.14)
(288, 21.77)
(296, 22.20)
(304, 21.88)
(312, 22.03)
(320, 21.91)
(328, 22.03)
(336, 22.27)
(344, 21.33)
(352, 21.30)
};
\addlegendentry{AM-MobileNet1D}
\end{axis}
\end{tikzpicture}
}
\subfigure[MIT]{
\begin{tikzpicture}
\begin{axis}[
width=0.95\columnwidth,
axis lines = left,
xlabel = Epochs,
ylabel = {FER ($\%$)},
]

\addplot [
    color=red,
    line width=0.55mm,
]coordinates {
(0, 87.96)
(8, 57.01)
(16, 47.45)
(24, 46.17)
(32, 43.11)
(40, 40.67)
(48, 39.89)
(56, 40.46)
(64, 38.70)
(72, 38.57)
(80, 38.33)
(88, 37.75)
(96, 38.99)
(104, 37.40)
(112, 36.70)
(120, 36.30)
(128, 37.21)
(136, 38.51)
(144, 36.20)
(152, 35.52)
(160, 36.69)
(168, 36.03)
(176, 36.15)
(184, 36.99)
(192, 37.19)
(200, 37.00)
(208, 38.75)
(216, 35.80)
(224, 36.33)
(232, 35.20)
(240, 35.97)
(248, 34.75)
(256, 35.55)
(264, 36.73)
(272, 35.63)
(280, 35.79)
(288, 35.13)
(296, 38.29)
(304, 35.06)
(312, 35.39)
(320, 35.78)
(328, 35.75)
(336, 35.47)
(344, 35.26)
(352, 35.13)
};
\addlegendentry{SincNet}

\addplot [
color=red,
dashed,
line width=0.55mm,
]
coordinates {
(0, 89.15)
(8, 57.25)
(16, 47.67)
(24, 41.66)
(32, 38.29)
(40, 37.65)
(48, 36.09)
(56, 35.61)
(64, 34.30)
(72, 35.90)
(80, 33.23)
(88, 33.27)
(96, 33.15)
(104, 33.29)
(112, 32.89)
(120, 32.24)
(128, 33.80)
(136, 31.61)
(144, 32.81)
(152, 30.80)
(160, 30.54)
(168, 30.78)
(176, 32.67)
(184, 30.30)
(192, 29.89)
(200, 30.18)
(208, 32.23)
(216, 29.97)
(224, 30.43)
(232, 30.36)
(240, 29.86)
(248, 29.67)
(256, 31.06)
(264, 30.53)
(272, 29.79)
(280, 29.58)
(288, 29.97)
(296, 29.90)
(304, 29.63)
(312, 29.28)
(320, 29.88)
(328, 29.84)
(336, 29.36)
(344, 29.53)
(352, 29.50)
};
\addlegendentry{AM-SincNet}

\addplot [
    color=blue,
    line width=0.55mm,
]coordinates {
(0, 81.32)
(8, 48.41)
(16, 42.33)
(24, 43.36)
(32, 43.31)
(40, 41.71)
(48, 40.63)
(56, 41.78)
(64, 42.48)
(72, 40.78)
(80, 41.19)
(88, 40.89)
(96, 40.89)
(104, 40.56)
(112, 40.56)
(120, 40.71)
(128, 40.43)
(136, 40.04)
(144, 40.21)
(152, 39.98)
(160, 39.85)
(168, 39.51)
(176, 40.34)
(184, 39.24)
(192, 40.01)
(200, 40.22)
(208, 39.51)
(216, 39.51)
(224, 40.10)
(232, 39.53)
(240, 40.07)
(248, 39.60)
(256, 40.15)
(264, 40.63)
(272, 39.35)
(280, 39.09)
(288, 39.79)
(296, 38.84)
(304, 39.11)
(312, 38.84)
(320, 39.72)
(328, 38.81)
(336, 40.23)
(344, 39.26)
(352, 38.75)
};
\addlegendentry{MobileNet1D}

\addplot [
    color=blue,
    dashed,
    line width=0.55mm,
]coordinates {
(0, 80.89)
(8, 46.94)
(16, 39.83)
(24, 39.68)
(32, 39.76)
(40, 39.58)
(48, 39.89)
(56, 38.85)
(64, 37.93)
(72, 38.30)
(80, 37.42)
(88, 37.51)
(96, 37.54)
(104, 37.24)
(112, 36.95)
(120, 36.86)
(128, 37.07)
(136, 36.78)
(144, 37.17)
(152, 36.54)
(160, 36.85)
(168, 36.48)
(176, 36.61)
(184, 36.52)
(192, 36.45)
(200, 36.66)
(208, 36.34)
(216, 36.52)
(224, 35.90)
(232, 35.97)
(240, 35.55)
(248, 36.33)
(256, 36.63)
(264, 36.25)
(272, 35.89)
(280, 36.27)
(288, 35.66)
(296, 35.42)
(304, 35.79)
(312, 35.72)
(320, 35.59)
(328, 35.82)
(336, 35.50)
(344, 35.67)
(352, 35.30)
};
\addlegendentry{AM-MobileNet1D}

\end{axis}
\end{tikzpicture}
}
\caption{Frame Error Rate (FER) over the training epochs.}
\label{fig:fer_timit_mit}
\end{figure*}
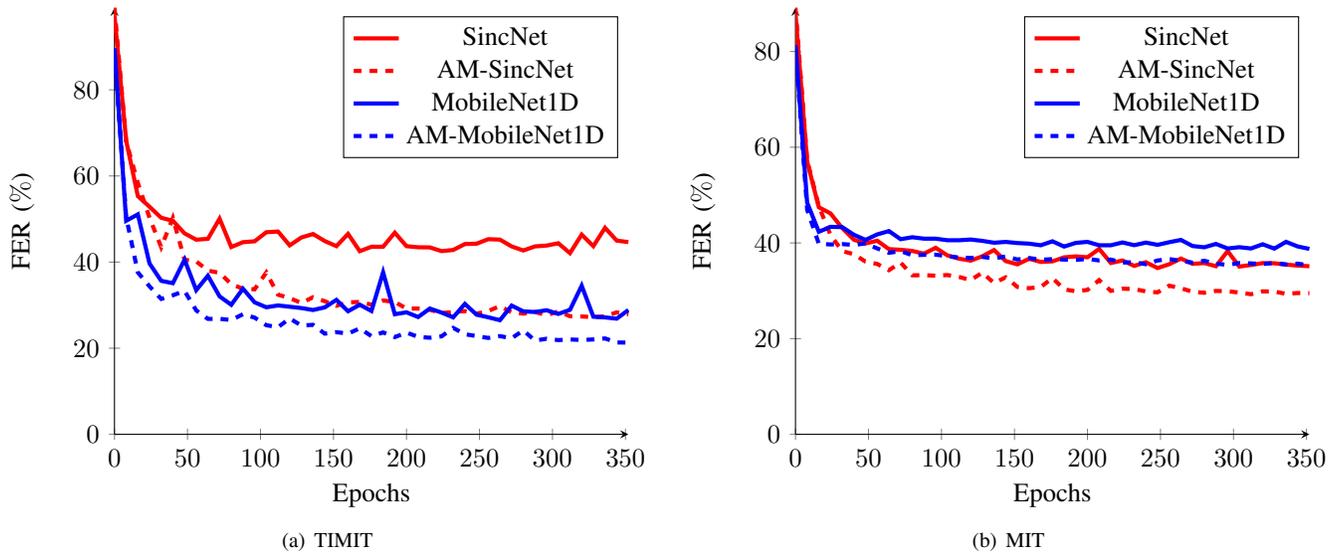

\subsection{Training}
\label{training}

The pre-processing step on the audio samples are the same used in \cite{SincNet} and \cite{AMSincNet}, where a normalization was applied to the audio signals. Also, for the TIMIT dataset, the non-speech interval was removed from the beginning and the end of the sentences. Moreover, the waveform of each audio sample was divided into $200$ms frames with $10$ms overlap; these frames were used to feed the network.

For training purposes, every network model was configured to use the RMSprop optimizer. Also, learning rate was set to $lr\!=\!0.001$, with $\alpha\!=\!0.95$, and $\epsilon\!=\!10^{-7}$. We set mini-batches to be at size $128$. The AM-Softmax comes with two more parameters compared to the traditional Softmax, the parameter $s$ for scaling the features and the parameter $m$ to control the size of the additive margin. According to \cite{AM-Softmax}, the scaling factor could be learned by the network, but it would take too long to optimize its value. Also, \cite{AM-Softmax} states that the scaling factor $s$ needs to be high enough in order to work well. Thus the author suggests setting $s$ to a fixed value of $s=30$ to speed up the training process. On the other hand, for the $m$ parameter, \cite{AMSincNet} made several experiments with the AM-SincNet model in the TIMIT dataset to see the influence of the margin parameter on the FER, according to the authors every tested value of $m$ converges the FER to the same place, thus in our experiments with the AM-MobileNet1D we set $m=0.5$. We also have added an $epsilon$ constant of value $10^{-11}$ to the AM-Softmax equation in order to avoid a division by zero in the required places.

Each model was trained for exactly $360$, which shows to be enough to exploit the model's efficiency. We also set a $seed = 1234$ to maintain the consistency of the experiments. The experiments run in a Core i5-9400F machine with a NVIDIA RTX 2060 GPU. We use CUDA 10.1 and Pytorch 1.0. 

\subsection{Metrics}
\label{metrics}

To evaluate the proposed models and compare them to the baseline approaches, we used the Frame Error Rate (FER) and Classification Error Rate (CER), two widely used metrics on the literature for the Speaker Recognition problem. We also compare the models by their inference time, which is calculated using the average time in milliseconds that the machine takes to process $6,561$ batches of size $128$, their number of parameters, and the model size in megabytes.

The performed experiments may be reproduced by using the code that we made available online at the GitHub \footnote{https://github.com/joaoantoniocn/AM-MobileNet1D} \footnote{https://github.com/joaoantoniocn/AM-SincNet}.

\section{Results}
\label{results}

In this section, we explore the results of the experiments. The proposed methods show to be competitive in terms of FER and CER. The methods obtained an equivalent or even better results than the SincNet and AM-SincNet models on the TIMIT and MIT dataset. Additionally, the proposed methods, MobileNet1D and AM-MobileNet1D, are about $7$ times faster, have $8$ times fewer parameters, and take up to $8$ times less space on disk than the SincNet and AM-SincNet.%

Table \ref{tab:results_error} shows the results of the experiments on TIMIT and MIT dataset using the metrics Frame Error Rate (FER) and Classification Error Rate (CER). The best result in each dataset is highlighted in bold. Based on these results, it is possible to see that in every scenario, the AM-Softmax loss function improved the performance of the models. Indeed, the benefit of using the AM-Softmax loss function on the Speaker Recognition models is quite clear, mainly on the SincNet model for the TIMIT dataset. The FER and CER calculated on the TIMIT dataset for the proposed method MobileNet1D are about $40\%$ and $50\%$ smaller (better) than the ones measured from SincNet, respectively. Also, when comparing the MobileNet1D with the AM-SincNet in the same dataset, they are practically equivalent, the MobileNet1D got a slightly better FER, while the AM-SincNet got a slightly better CER. For this dataset, the model with the best results was the proposed AM-MobileNet1D, it got the lowest (best) FER and CER over the models. According to Table \ref{tab:results_error}, the FER measured on the TIMIT dataset for the AM-MobileNet1D was $19\%$ better than the MobileNet1D, $23\%$ better than the AM-SincNet, and $52\%$ better than the traditional SincNet. In the same way, in terms of CER it was $24\%$ better than the MobileNet1D, $14\%$ better than the AM-SincNet, and $62\%$ better than the SincNet. On the other hand, the SincNet and the AM-SincNet overcome the proposed models in terms of FER and CER for the MIT dataset. Still, the proposed models are quite good on this dataset showing an accuracy of about $98\%$.

Table \ref{tab:results_error} also shows the results of inference time in milliseconds, the number of parameters, and the model size (in megabytes) for each model. The inference time results are in the format $time \pm standard$ $deviation$. The results show that the proposed models are about $7$ times faster, have $8$ times fewer parameters, and take up to $8$ times less space on disk than the SincNet and the AM-SincNet. The proposed MobileNet1D and AM-MobileNet1D take only $11.6$ megabytes on disk, against $91.2$mb from the SincNet and AM-SincNet. In fact, these results show that the proposed models are suitable for being embedded on Mobile devices, showing high performance and low inference latency for the speaker recognition problem.

Figures \ref{fig:fer_timit_mit} (a) and (b) show the evolution of the Frame Error Rate (FER) over the training epochs on the TIMIT and MIT dataset, respectively. There, the SincNet and AM-SincNet are plotted in red, while the MobileNet1D and AM-MobileNet1D are in blue. Also, the methods using the AM-Softmax loss function are dashed. On both figures, the methods using the AM-Softmax loss function perform better than the ones using the traditional Softmax. From Figure \ref{fig:fer_timit_mit} (a) it is clear the difference on FER from both proposed models, MobileNet1D and AM-MobileNet1D, to the traditional SincNet. Also, the MobileNet1D and the AM-SincNet are equivalent in terms of FER on the TIMIT dataset. Figure \ref{fig:fer_timit_mit} (b) shows the results for the MIT dataset. There we can see that unlike Figure \ref{fig:fer_timit_mit} (a), the results are closer to each other, showing a modest difference between them. Yet, the AM-SincNet has performed better on this dataset, while the MobileNet1D and the SincNet presented almost the same result.

In general, the proposed AM-MobileNet1D show high performance in the TIMIT dataset (the best of all models), while got less than $1\%$ increase in the CER on the MIT dataset compared to the SincNet. Indeed, the great advantages of the proposed MobileNet1D and AM-MobileNet1D are its inference time, which is about $7$ times faster than the SincNet and the AM-SincNet, and its model size which takes only $11.6$ megabytes on disk. Thus, making it an excellent choice to tackle the speaker recognition problem, especially in environments where the system needs a quick or embedded answer.

\section{Conclusion} 
\label{conclusion}

This paper has proposed new methods for directly processing waveform audio on mobile devices for the speaker recognition task. The proposed methods MobileNet1D and AM-MobileNet1D got about $7$ times faster results with a low error rate compared to the hybrid methods SincNet and AM-SincNet. The AM-MobileNet1D also got the best Frame Error Rate (FER) and Classification Error Rate (CER) for the TIMIT dataset. Another advantage of the proposed methods lies in their model size, which takes only $11.6$ megabytes on the disk.

For future work, we should consider larger datasets like VoxCeleb1 \cite{voxceleb1} and VoxCeleb2 \cite{voxceleb2}. VoxCeleb1 contains over 100k samples distributed in about 1.2k celebrities, and the samples were extracted from videos uploaded on YouTube. In the same way, VoxCeleb2 \cite{voxceleb2} has over a million samples from over 6k speakers. Thus, it would be good to see how the MobileNet1D and the AM-MobileNet1D behaves on larger datasets.

\bibliographystyle{bibtex/bst/IEEEtran}
\bibliography{bibtex/bib/IEEEexample}

\end{document}